# Axial acoustic radiation force on rigid oblate and prolate spheroids in Bessel vortex beams of progressive, standing and quasi-standing waves

F.G. Mitri

A B S T R A C T

The analysis using the partial-wave series expansion (PWSE) method in spherical coordinates is extended to evaluate the acoustic radiation force experienced by rigid oblate and prolate spheroids centered on the axis of wave propagation of high-order Bessel vortex beams composed of progressive, standing and quasi-standing waves, respectively. A coupled system of linear equations is derived after applying the Neumann boundary condition for an immovable surface in a non-viscous fluid, and solved numerically by matrix inversion after performing a single numerical integration procedure. The system of linear equations depends on the partial-wave index $n$ and the order of the Bessel vortex beam $m$ using truncated but converging PWSEs in the least-squares sense. Numerical results for the radiation force function, which is the radiation force per unit energy density and unit cross-sectional surface, are computed with particular emphasis on the amplitude ratio describing the transition from the progressive to the pure standing waves cases, the aspect ratio (i.e., the ratio of the major axis over the minor axis of the spheroid), the half-cone angle and order of the Bessel vortex beam, as well as the dimensionless size parameter. A generalized expression for the radiation force function is derived for cases encompassing the progressive, standing and quasi-standing waves of Bessel vortex beams. This expression can be reduced to other types of beams/waves such as the zeroth-order Bessel non-vortex beam or the infinite plane wave case by appropriate selection of the beam parameters. The results for progressive waves reveal a tractor beam behavior, characterized by the emergence of an attractive pulling force acting in opposite direction of wave propagation. Moreover, the transition to the quasi-standing and pure standing wave cases shows the acoustical tweezers behavior in dual-beam Bessel vortex beams. Applications in acoustic levitation, particle manipulation and acousto-fluidics would benefit from the results of the present investigation.

## 1. Introduction

Radiation force simulations in acoustical tweezers applications have become an indispensable tool to scientific research dealing with the interaction of acoustical waves (or beams) with objects, especially in acoustofluidics [1, 2], particle manipulation [3] and acoustical tweezers [4-17], elasticity imaging [18], and acoustic levitation [19-24] to name a few applications. Numerical computations are essential as they provide guidance for optimal experimental design purposes. Furthermore, since experiments require adequate instrumentation and hardware equipment (which may be often expensive), and are time-consuming so in practice only a limited number can be performed on well-defined geometries, most investigations resort initially to numerical computations of the radiation force of acoustical waves exerted on an object placed along their path. By developing fast and accurate computational tools, precise radiation force modeling of any scenario of interest can be made possible. This includes extreme cases satisfying certain limits which may not be entirely attainable experimentally; for example, the completely rigid or soft particle cases. Clearly, there is a continuing need for improved modeling, analytical theories and computer simulations, especially when more complex (non-circular) geometries and/or beam profiles of arbitrary wavefronts are considered.

Various investigations limited to the long wavelength limit examined the acoustic radiation force on disks [25-28] and prolate spheroids [29] in plane progressive and standing waves. Moreover, the finite-element method [30] (FEM) has been utilized by means of the shape perturbation method to evaluate the acoustic radiation force on a (non-spherical) rigid spheroid in *plane* standing waves.

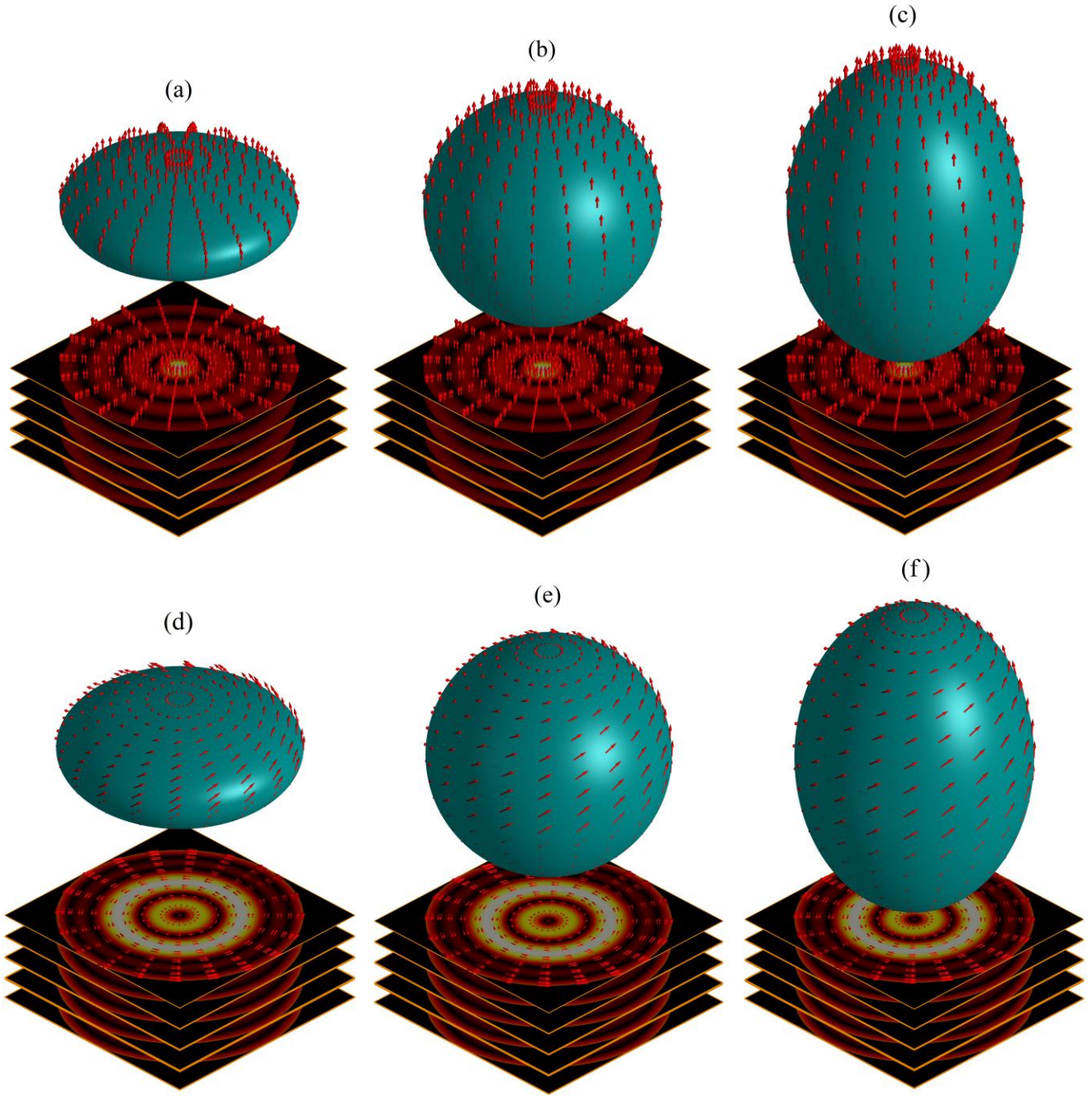

**Fig. 1.** This figure shows the computational plots (with stereographic projections in the bottom plane superimposed on the cross-sectional profile of the beam) corresponding to the spatial distribution of the intensity vector field of a zero-order Bessel (non-vortex) beam [panels (a)-(c)], and a first-order Bessel spiral (or vortex) beam ($m = 1$) [panel (d)-(f)] illuminating spheroidal particles with different aspect ratios centered on the axis of the incident beam. The particle in panels (a),(d) has the shape of an oblate spheroid, whereas in panels (b),(e), the particle is a sphere. In panels (c),(f), the particle has the shape of a prolate spheroid. The characteristic of the vortex is clearly shown in panels (d)-(f) with an intensity null at the center of the beam. Depending on the vortex helicity ($m = \pm 1$), the arrows representing the intensity vector field can be directed counter-clockwise or clockwise, respectively.

Notice, however, that the shape perturbation method is only applicable to a near-spherical particle, and leads to significant errors for the cases of moderately to highly elongated spheroids. An analysis based on the boundary element method (BEM) [31] has been also developed, in which the acoustic radiation force experienced by non-spherical particles has been computed for cases where the target's dimensions are much smaller than the wavelength of the incident illuminating plane waves (i.e. Rayleigh limit). The lack of adequate data beyond the long-wavelength limit as well as the restriction to the infinite plane wave case provided the impetus to further develop an improved solution [32] based on the modal-matching method in spherical coordinates to compute the acoustic radiation force on rigid oblate and prolate spheroids applicable to zeroth-order Bessel beams, including the infinite plane wave case.

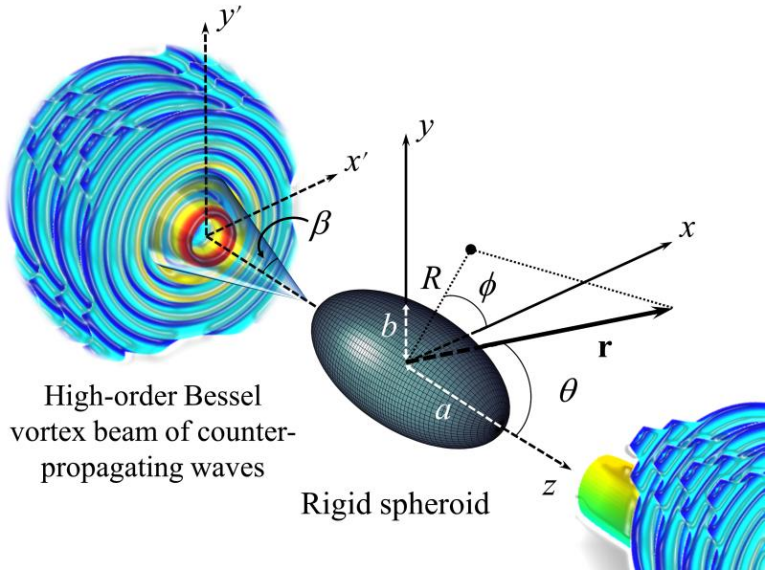

**Fig. 2.** The schematic describing the interaction of a monochromatic high-order Bessel vortex "hollow" beam composed of counter-propagating waves and centered on the major axis $a$ of a rigid fixed (sound impenetrable) prolate spheroid at end-on incidence. The parameters $\theta$ and $\phi$ are the polar and azimuthal angles, respectively, in a spherical coordinates system.

In contrast to plane (or Gaussian) waves, Bessel beams possess several features, such as the ability to reform after encountering an obstruction [33, 34]. They also have a limited-diffraction capability such that they maintain a relatively long depth-of-field as they propagate in space [35]. Due to these advantages, such beams are well established both theoretically and experimentally, and related innovative applications in fundamental and applied physics are increasingly expanding in various fields.

The zeroth-order Bessel beam is of non-vortex type, while the higher-order is of vortex (i.e. spiral or helicoidal) type, such that its incident velocity potential (or pressure) field varies according to $\exp(i|m|\phi)$, where $m$ is any (real) integer number known as the order or the topological charge, and $\phi$ is the azimuthal angle. This effect can be clearly emphasized by considering a computational example for the intensity vector field $\mathbf{I} = \Re\{p^*\mathbf{v}\}$ of a Bessel vortex beam, where the parameters $p$ and $\mathbf{v}$ denote the incident linear pressure and vector velocity, respectively. The superscript * denotes the conjugate of a complex number. The results are displayed in Fig. 1 for the computational plots (with stereographic projections in the bottom plane superimposed on the cross-sectional profile of the beam) corresponding to the spatial distribution of the intensity vector field (shown by the arrows) of a Bessel beam sampled uniformly on the surface of the spheroid. The plots in panels (a)-(c) correspond to a zeroth-order Bessel (non-vortex) beam where a maximum intensity at the center of the beam is produced, whereas panels (d)-(f) are for a first-order Bessel spiral (or vortex) beam with a unit topological charge (i.e., $m = 1$) where a null [36] (or phase singularity along the axis of wave propagation) is manifested at the center of the beam. The incident Bessel waves illuminate spheroidal particles with different aspect ratios centered on the axis of the beam. The particle in panels (a),(d) has the shape of an oblate spheroid, whereas in panels (b),(e), the particle is a sphere. In panels (c),(f), the particle has the shape of a prolate spheroid. The characteristic of the vortex is clearly shown in panels (d)-(f) with an intensity null at the center of the beam. Depending on the vortex helicity ($m = \pm 1$), the arrows representing the intensity vector field can be directed counter-clockwise or clockwise, respectively.

While there exists significant literature on the interaction of acoustical Bessel beams with spherical objects including several investigations focused on the (arbitrary) scattering[37-39] and radiation forces[40-44], the spheroidal object has been only recently considered [32, 45-49]. As the properties of the higher-order beam solutions (using progressive waves) [48] differ considerably from the fundamental (zeroth-order) case [32], it is of some importance to develop an appropriate analytical formalism to analyze and compute numerically the acoustic radiation force of Bessel vortex beams of standing and quasi-standing waves exerted on spheroids. Typically, standing and quasi-standing waves are obtained by counter-propagating two (or more) Bessel vortex beams. This configuration is also known as the dual-beam tweezers as suggested in the original version of acoustical tweezers [50]. Note that the formalism used for Bessel vortex beams of progressive waves [48] is not applicable to the dual-beam configuration, nor it can be considered as an approximate solution because the localizing force strengths for the representative standing and quasi-standing wave modes depend on the spatial phase which is not the case for progressive waves. Moreover, the standing and quasi-standing waves in the dual-beam configuration form stable positioning in acoustical potentials, a situation that can be hardly achieved using the single-beam of progressive waves. Such limitations provide the motivation and impetus to undertake the present analysis, and develop an analytical method which encompasses all types of waves in Bessel vortex beams, ranging from progressive, standing to quasi-standing waves.

In this analysis, the acoustic scattering of Bessel vortex beams of quasi-standing waves by a rigid (sound impenetrable) oblate or prolate spheroid (Fig. 2) is first solved first using the partial-wave series expansion (PWSE) method in spherical coordinates. Then, it is used to derive a generalized analytical expression for the axial acoustic radiation force (i.e., acting along the direction of wave propagation) that is applicable to the cases of progressive, quasi-standing and standing waves in Bessel vortex beams. It is important to note here that such spheroidal (convex-like) surfaces present a serious challenge because the method of separation of variables (used to evaluate the scattering and subsequently the radiation force) becomes inapplicable. In other words, the spherical-wave functions used in the method of separation of variables become non-orthogonal on the object's surface, consequently, adequate convergence and accuracy of the results can be hardly achieved. Nonetheless, this difficulty is resolved by using the PWSE using an improved methodology based on modal matching. The method requires solving a system of linear equations by matrix inversion procedures [i.e. Eq.(15) in the following] which depends on the partial-wave index $n$ and the order $m$ of the Bessel vortex beam. For the case considered in the present manuscript, the Neumann boundary condition for a rigid immovable surface is satisfied in the least-squares sense with negligible truncation numerical error. This original semi-analytical approach developed for Bessel vortex beams is demonstrated for finite oblate and prolate spheroids, where the mathematical functions describing the spheroidal geometry are written in a form involving single angular (polar) integrals that are numerically computed using Gauss-Legendre quadratures. Then, the axial radiation force is evaluated stemming from an analysis of the far-field scattering, with particular emphasis on the aspect ratio (i.e., the ratio of the major axis over the minor axis of the spheroid), the half-cone angle $\beta$ and the order $m$ of the Bessel vortex beam, as well as the dimensionless size parameter $kb$. Moreover, the radiation force function expression reduces to progressive or equi-amplitude standing waves depending on the choice of the coefficient $R$, defined as an amplitude-ratio factor of the waves (Section 2).

## 2. Theoretical formalism

Consider an acoustical monochromatic high-order Bessel vortex beam of order $|m|$ propagating in a nonviscous fluid, and incident upon a spheroid centered on its axis of wave propagation [i.e., end-on incidence (Fig. 2)]. The spheroid has an equatorial radius $b$ (known as the semi-minor axis), and $a$ is the distance from the center to the pole along the symmetry axis $z$, corresponding to the semi-major axis.

The incident field is assumed to be composed of two counter-propagating Bessel vortex beams of the same order but with different amplitudes. This generally defines an acoustic velocity potential field of quasi-standing waves.

In a system of spherical coordinates $(r, \theta, \phi)$ with its origin chosen at the center of the spheroid, the incident velocity potential field is expressed as a partial-wave series expansion (PWSE) as [51],

$$\begin{aligned}\Phi_i &= e^{-i\omega t}\left[\varphi_0 e^{ik_z(z+h)} + \varphi_1 e^{-ik_z(z+h)}\right] J_{|m|}(k_r r \sin\theta) e^{i|m|\phi} \\ &= \varphi_0 e^{i(|m|\phi-\omega t)} \sum_{n=|m|}^{\infty} i^{(n-m)} \psi_n (2n+1) \frac{(n-m)!}{(n+m)!} j_n(kr) P_n^m(\cos\theta) P_n^m(\cos\beta),\end{aligned} \quad (1)$$

where $\psi_n = \left[e^{ik_z h} + R(-1)^n e^{-ik_z h}\right]$, $R = \varphi_1/\varphi_0$, with the assumption that $|\varphi_0| > |\varphi_1|$ so that $0 \leq |R| \leq 1$. The parameters $k_z = k\cos\beta$ and $k_r = k\sin\beta$ are the axial and radial wave-numbers, $k = \sqrt{k_z^2 + k_r^2} = \omega/c = 2\pi/\lambda$, $k$ is the wave number, $\omega$ is the angular frequency, $c$ is the speed of sound in the fluid medium with a density denoted by $\rho$, the parameter $\lambda$ being the wavelength, $\beta$ is the half-cone angle formed by the wave-number $k$ relative the axis of wave propagation along the $z$-direction (See Fig. 2), $h$ is the distance in the $z$-direction from the center of the spheroid to the nearest velocity potential antinode, the angle $\theta$ is the scattering polar angle relative to the beam axis of wave propagation $z$, $\phi$ is the azimuthal angle in the transverse $(x,y)$ plane, the function $J_{|m|}(.)$ is the cylindrical Bessel function of the first kind of order $|m|$, $r$ is the distance from the center of the coordinate system to an observation point, $j_n(.)$ is the spherical Bessel function of order $n$ and $P_n^m(.)$ are the associated Legendre functions.

The scattered velocity potential field from the spheroid is expressed as,

$$\Phi_s = \varphi_0 e^{i(|m|\phi-\omega t)} \sum_{n=|m|}^{\infty} i^{(n-m)} \psi_n (2n+1) \frac{(n-m)!}{(n+m)!} S_n h_n^{(1)}(kr) P_n^m(\cos\theta) P_n^m(\cos\beta), \quad (2)$$

where $S_n$ are the spheroid's scattering coefficient to be determined by imposing the Neumann's boundary condition in a non-viscous fluid, and $h_n^{(1)}(.)$ is the spherical Hankel function of the first kind.

It is important to note here that the spherical Hankel functions used in Eq.(2) to describe the scattered waves from a non-spherical object in the surrounding fluid medium, form a complete set [52] everywhere outside a minimum spherical surface of radius $r_0 = \max(a,b)$ circumscribing the spheroid. This fact may cause (in general) an apparent complication restricting the application of any boundary condition directly on the surface of the object. Nevertheless, if the surface is free from any singularity, or in other words, if the surface shape function of the object is expressed by means of continuous regular functions, the boundary condition will be satisfied at the surface of the object owing to Huygens' principle [52]. Note however that this process of analytic continuation may not be applicable to highly elongated spheroids or at high frequencies because adequate convergence can be hardly achieved. There exist several methods and techniques that could be used to advantage for highly elongated (or extremely flat) surfaces, such as the iterative Gram-Schmidt orthogonalization procedure [53], a boundary condition enforcement by point-matching [54], the use of extended-precision floating-point variables [55], or other methods.

For a spheroid, the surface shape function $S_\theta$ only depends on the polar angle $\theta$. Its expression is given by,

$$S_\theta = \left[\left(\cos\theta/a\right)^2 + \left(\sin\theta/b\right)^2\right]^{-1/2}. \quad (3)$$

Imposing the Neumann boundary condition for the total (incident + scattered) steady-state (time-independent) velocity potential field for a rigid immovable spheroid at $r = S_\theta$, leads to the following equation

$$\nabla\left(\Phi_i + \Phi_s\right)\cdot\mathbf{n}\Big|_{r=S_\theta} = 0, \quad (4)$$

where,

$$\mathbf{n} = \mathbf{e}_r - \left(\frac{1}{S_\theta}\right)\frac{dS_\theta}{d\theta}\mathbf{e}_\theta, \quad (5)$$

with $\mathbf{e}_r$ and $\mathbf{e}_\theta$ denoting the unit vectors along the radial and polar directions, respectively.

Substituting Eqs.(1) and (2) into Eq.(4) using Eq.(5), a system of linear equations is obtained as,

$$\sum_{n=|m|}^{\infty} i^n \psi_n (2n+1)\frac{(n-m)!}{(n+m)!} P_n^m(\cos\beta)\left[\Gamma_n^m(\theta) + S_n\Lambda_n^m(\theta)\right] = 0, \quad (6)$$

where the functions $\Gamma_n^m(\theta)$ and $\Lambda_n^m(\theta)$ are expressed, respectively, as

$$\begin{Bmatrix}\Gamma_n^m(\theta)\\ \Lambda_n^m(\theta)\end{Bmatrix} = k\begin{Bmatrix}j_n'(kS_\theta)\\ h_n^{(1)'}(kS_\theta)\end{Bmatrix} P_n^m(\cos\theta) - \begin{Bmatrix}j_n(kS_\theta)\\ h_n^{(1)}(kS_\theta)\end{Bmatrix}\frac{1}{S_\theta^2}\frac{dS_\theta}{d\theta}\frac{dP_n^m(\cos\theta)}{d\theta}, \quad (7)$$

where the primes denote derivation with respect to the argument. For a fixed partial-wave index $n$, order $m$ of the Bessel beam and wave number $k$, the angular functions $\Gamma_n^m(\theta)$ and $\Lambda_n^m(\theta)$ depend only on the variable angle $\theta$ since the surface shape function given by Eq.(3) has azimuthal symmetry. To solve Eq.(6), the angular dependency must be eliminated in Eq.(7). This is done by equating Eq.(6) to a PWSE series with appropriate functions as,

$$\sum_{n=|m|}^{\infty} i^n \psi_n (2n+1)\frac{(n-m)!}{(n+m)!} P_n^m(\cos\beta)\left[\Gamma_n^m(\theta) + S_n\Lambda_n^m(\theta)\right]$$
$$= \sum_{n=|m|}^{\infty}\left[\Upsilon_n^m + S_n\Omega_n^m\right]P_n^m(\cos\theta) \quad (8)$$
$$= 0,$$

where $\Upsilon_n^m$ and $\Omega_n^m$ are the coefficients [independent of the angle $\theta$] to be determined after applying to Eq.(8) the orthogonality condition of the associated Legendre functions [56],

$$\int_0^\pi P_n^m(\cos\theta) P_\ell^m(\cos\theta) \sin\theta d\theta = \frac{2}{(2\ell+1)} \frac{(\ell+m)!}{(\ell-m)!} \delta_{n,\ell}, \quad (9)$$

where $\delta_{n,\ell}$ is the Kronecker delta function.

Subsequently, equating the left- and right-hand sides in Eq.(8) for each pair set $(n, m)$, a new system of linear equations satisfying the Neumann boundary condition is obtained, which allows appropriate determination of the scattering coefficients $S_n$ for the spheroid. The new system of linear equations to be solved is now rewritten as,

$$\sum_{\ell=|m|}^\infty \left[ \Upsilon_\ell^m + S_n \Omega_\ell^m \right] = 0, \quad (10)$$

where,

$$\begin{Bmatrix} \Upsilon_\ell^m \\ \Omega_\ell^m \end{Bmatrix} = \sum_{n=|m|}^\infty i^n \psi_n (2n+1) \frac{(n-m)!}{(n+m)!} P_n^m(\cos\beta) \int_0^\pi \begin{Bmatrix} \Gamma_n^m(\theta) \\ \Lambda_n^m(\theta) \end{Bmatrix} P_\ell^m(\cos\theta) \sin\theta d\theta. \quad (11)$$

Notice that Eq.(11) explicitly shows that the coefficients $\Upsilon_\ell^m$ and $\Omega_\ell^m$ are the result of the coupling between the coefficients characterizing the incident beam with the integration of functions describing the geometrical shape of the object. The integrals are evaluated numerically before the summation is carried out, and the linear system of equations [given by Eq.(10)] can be solved by appropriate truncation of the series and subsequent matrix inversion.

In general, Eq.(10) is satisfied when the maximum partial wave index $\ell_{max} \to \infty$. In practice however, infinite summations cannot be performed, and therefore the series must be truncated at some point. It is important to know where to truncate the series so as to minimize the truncation error. For separable geometries (such as a sphere or cylinder) the wave-functions functions are orthogonal and convergence is reached rapidly. For a non-separable geometry, the wave-functions are no longer orthogonal and a linear system of equations has to be adequately truncated and solved to evaluate the scattering coefficients of the object. Since all of the coefficients are coupled to one another, truncation of the series will affect the accuracy of $S_n$ if a convergence criterion is not properly satisfied. The truncation limits $\ell_{max} = n_{max}$ have been chosen here such that $\left| S_{n+n_{max}} / S_m \right| \sim 10^{-16}$, for $(n,\ell) = m, \ldots, (n_{max}, \ell_{max})$, which yielded adequate convergence and negligible truncation error[47]. This criterion in practice requires that $(n_{max}, \ell_{max})$ be large [47]. For instance, such convergence has been previously attained for oblate and prolate spheroid with an aspect ratio of 3:1 [32, 46]. This also applies for the case where the frequency of the incident beam increases, as more terms in the series are needed to ensure convergence. However, one may not be able to continually increase the truncation limits as this procedure comes with a substantial computational cost. Moreover, with the increasing number of terms, the computational accuracy of the spherical Hankel functions begins to degrade rapidly, and creates an eventual ill-conditioning [57, 58] while solving the linear system given by Eq.(10) by matrix inversion procedures. Moreover, the numerical computations have considered spheroids with a maximum aspect ratio not exceeding 3:1, corresponding to cases that can be adequately treated using the present method.

Once the scattering coefficients $S_n$ are numerically evaluated, calculation of the acoustic radiation force becomes possible, stemming from an analysis based on the far-field scattering [32] from the spheroid. The method requires integrating the radiation stress tensor on a spherical surface (designated as $S_r$) located at a large distance, which encloses the spheroid.

The expression for the force is given by [13, 59]

$$\langle \mathbf{F} \rangle = \frac{1}{2} \rho k^2 \iint_{S_r} \Re \left\{ \left( \frac{i}{k} \partial_r \tilde{\Phi}_i - \tilde{\Phi}_i \right) \tilde{\Phi}_s^* - \left| \tilde{\Phi}_s \right|^2 \right\} d\mathbf{S}, \quad (12)$$

where $\tilde{\Phi}_i$ and $\tilde{\Phi}_s$ are the far-field ($kr \to \infty$) limits of Eq.(1) and Eq.(2), respectively, obtained by taking the large arguments limits of the spherical Bessel function $j_n(kr) \underset{kr \to \infty}{\approx} (kr)^{-1} \sin(kr - n\pi/2)$ and Hankel function of the first kind, $h_n^{(1)}(kr) \underset{kr \to \infty}{\approx} (kr)^{-1} i^{-(n+1)} e^{ikr}$. The parameter $\rho$ is the mass density of the surrounding fluid, the operator $\partial_r = \partial/\partial r$, $d\mathbf{S} = \mathbf{e}_r dS$, where the differential surface is $dS = r^2 \sin\theta d\theta d\phi$, the symbol $\langle . \rangle$ denotes time-averaging, $\Re\{\cdot\}$ is the real part of a complex number, and the superscript * denotes the conjugate of a complex function.

After some algebraic manipulation using the properties of the angular integrals (given explicitly in the Appendix of Ref.[[13]]), the axial component $F_z$ of the force is expressed as,

$$F_z = \langle \mathbf{F} \rangle \cdot \mathbf{e}_z = Y_{qst,z} S_c E_0, \qquad (13)$$

where $\mathbf{e}_z$ is the unit vector along the axial $z$-direction, $E_0 = \rho k^2 |\varphi_0|^2 / 2$, is a characteristic energy density, $S_c$ is a cross-sectional surface, and $Y_{J_m,z}^{qst}$ is the radiation force function, which is the radiation force per unit energy density and unit cross-sectional surface. Its expression is given by,

$$Y_{qst,z} = \frac{8\pi}{k^2 S_c} \sum_{n=|m|}^{\infty} \left\{ \begin{array}{l} \left[(-1)^{(n+m+1)} R \left[ \beta_n (1 + 2\alpha_{n+1}) - \beta_{n+1}(1 + 2\alpha_n) \right] \sin(2k_z h) \right] \\ -\frac{1}{2}(1 - R^2)\left[\alpha_n + \alpha_{n+1} + 2(\alpha_n \alpha_{n+1} + \beta_n \beta_{n+1})\right] \\ \times \frac{(n-m+1)!}{(n+m)!} P_n^m(\cos\beta) P_{n+1}^m(\cos\beta) \end{array} \right\}. \qquad (14)$$

where $\alpha_n = \Re\{S_n\}$, $\beta_n = \Im\{S_n\}$, and $\Im\{\cdot\}$ is the imaginary part of a complex number. Eq.(14) reduces to the radiation force function $Y_{p,z}$ for progressive waves (i.e., $R = 0$), or the radiation force function $Y_{st,z}$ for equi-amplitude "pure" standing waves (i.e., $R = 1$), depending on the value of $R$.

For a sphere, the mathematical expression given by Eq.(14) exactly agrees with previous results obtained for the radiation force [with $S_c = \pi a^2$, where $a$ is the sphere's radius] in high-order Bessel vortex beams of quasi-standing waves [60], progressive waves [40, 41] or standing waves [61], where in this case, $\alpha_n$ and $\beta_n$ indicate the real and imaginary parts of the scattering coefficients of the *sphere*. Furthermore, Eq.(14) can be applied to the zeroth-order Bessel beam incident on spheroids [32] by taking $m = 0$, and the infinite plane wave case by setting both $m$ and $\beta$ equal to zero.

3. **Numerical results and discussions**

The numerical determination of the acoustic radiation force function in Bessel vortex beams with emphasis on the transition from the progressive to the standing wave behavior requires the evaluation of the (yet unknown) scattering coefficients $S_n$. These coefficients are obtained by solving the truncated system of linear equations system

$$\sum_{\ell=|m|}^{\ell_{max}} \left[ \Upsilon_\ell^m + S_n \Omega_\ell^m \right] = 0. \qquad (15)$$

A MATLAB program code is developed to solve initially the system of linear equations given by Eq.(15), then evaluate the integrals in Eq.(11) using a standard numerical integration procedure based on the fast trapezoidal method with a sampling step of $\delta\theta = 10^{-5}$ which ensures adequate convergence and accuracy of the results. In the computations, $S_c = \pi b^2$, $kh = \pi/4$ (which is the correct parameter also used in Ref.[[32]], and not $k_z h = \pi/4$), and the non-dimensional size parameter $kb$ is varied in the range $0 < kb \le 7$.

The numerical verification and validation with previous results is performed for the cases of a zeroth-order Bessel beam [32] using a simplified form (i.e., $m = 0$) of the PWSE method presented here. For all the chosen examples (not shown here for brevity), the plots correlated exactly with those obtained previously in Ref.[[32]]. Note also that the PWSE method has been independently validated in the context of acoustic scattering theory [46] with the *T*-matrix formalism [45] and showed perfect agreement.

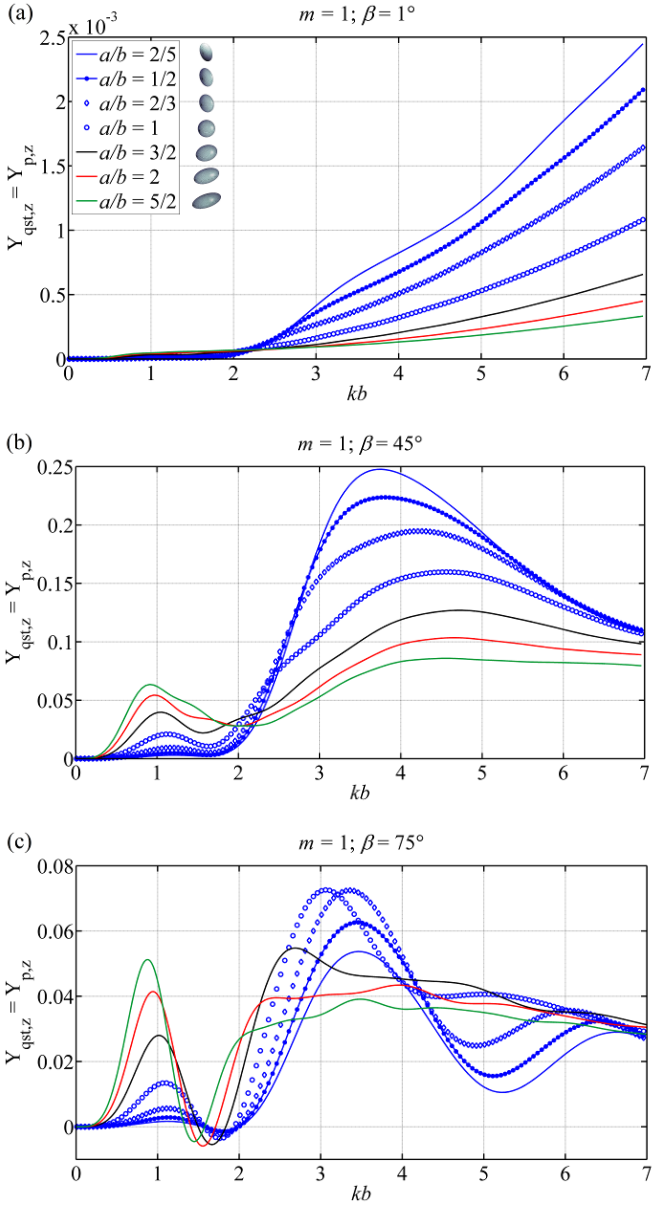
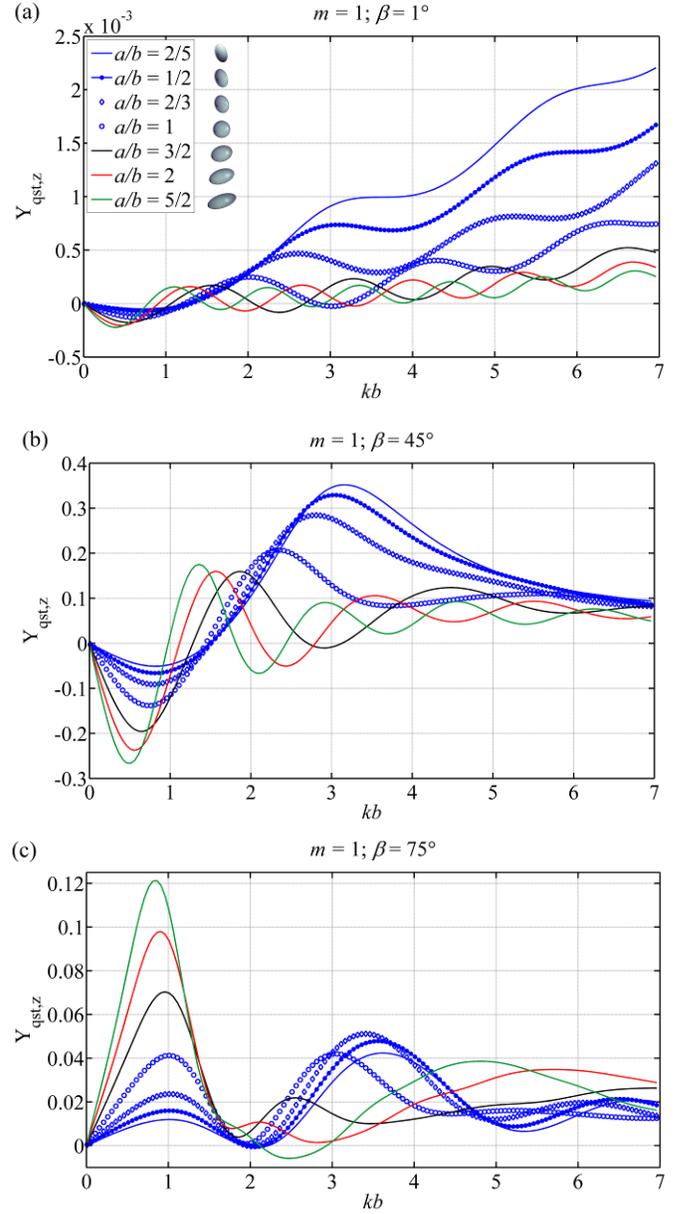

**Fig. 3.** Panels (a)-(c) show the effect of varying the half-cone angle $\beta$ of a first-order Bessel vortex beam of *progressive waves* (i.e. $R = 0$) on the radiation force function (14) for rigid immovable (fixed) spheroids with end-on incidence. The ratio $a/b$ is given on the top left side of panel (a), showing the transition from the oblate to the prolate spheroid.

**Fig. 4.** The same as in Fig. 3, but the Bessel vortex beam is composed of quasi-standing waves with $R = 0.5$.

Computations for the acoustic radiation force function given by Eq.(14) are now shown for three distinct half-cone angle values $\beta = 1°$, $45°$ and $75°$, corresponding to panels (a)-(c) of Fig. 3, respectively. The first-order Bessel vortex beam is composed of progressive waves (i.e., $R = 0$) centered on the spheroid with end-on incidence, as shown in Fig. 2. The effect of changing the spheroid geometry from the oblate to the prolate shape is also emphasized by varying the aspect ratio $a/b$ (which is the ratio of the major axis to the minor axis of the spheroid). When $a = b$, the spheroid corresponds to a sphere. At low $kb$ values ($< 2$), the radiation force function is larger the more prolate the spheroid becomes. This behavior changes as $kb$ increases ($> 2$) where the opposite behavior occurs; that is, the radiation force function decreases the more prolate the spheroid becomes, as shown in panels (a) and (b). Nevertheless, for larger half-cone angles, some exceptions occur; the radiation force function does not show a general similar behavior with the change in $kb$, as shown in panel (c). Moreover, there exists a $kb$-range ($1.25 \lesssim kb \lesssim 2$), over which the radiation force function for *progressive waves* is negative, which induces an attractive (instead of a repulsive) force on the spheroid (i.e. "tractor beam" [48] behavior for a spheroid). This phenomenon has been initially observed for a rigid sphere in Bessel vortex beams of any order (or topological charge) [40]. Notice also, that for a small half-cone angle $\beta$ (= 1°), the radiation force function is weaker as compared to the plots of panels (b) and (c). As $\beta$ increases, the central hollow (intensity null) region diameter decreases, and the spheroid is illuminated by a larger portion of the beam leading to the increase in the radiation force function plots.

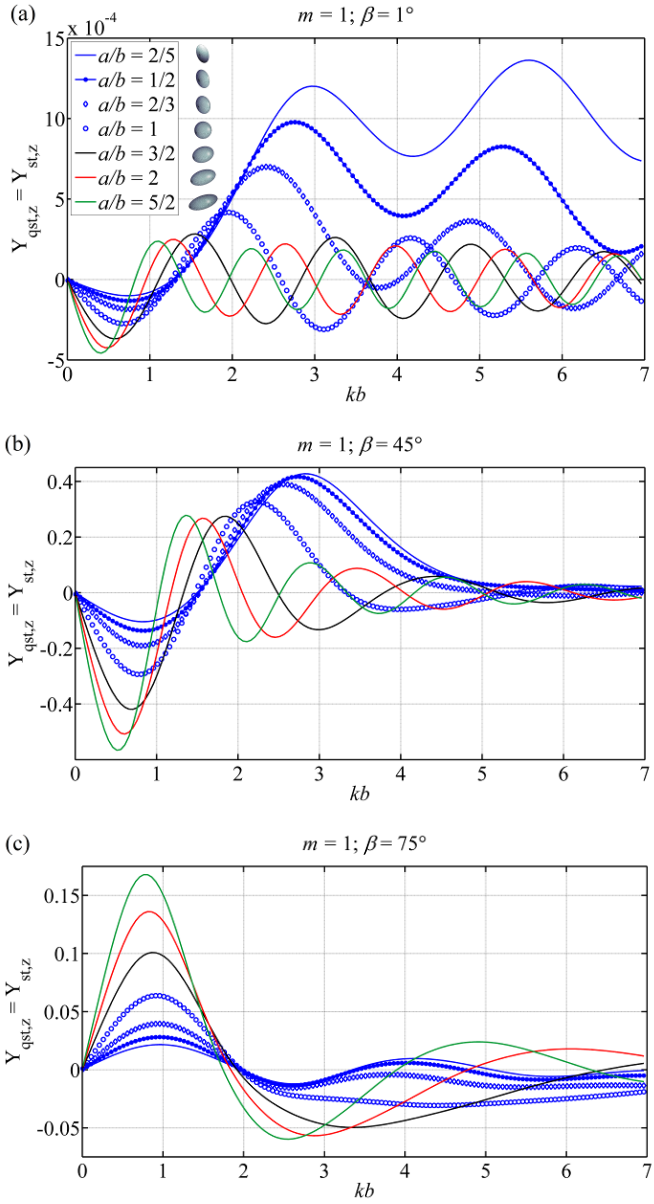
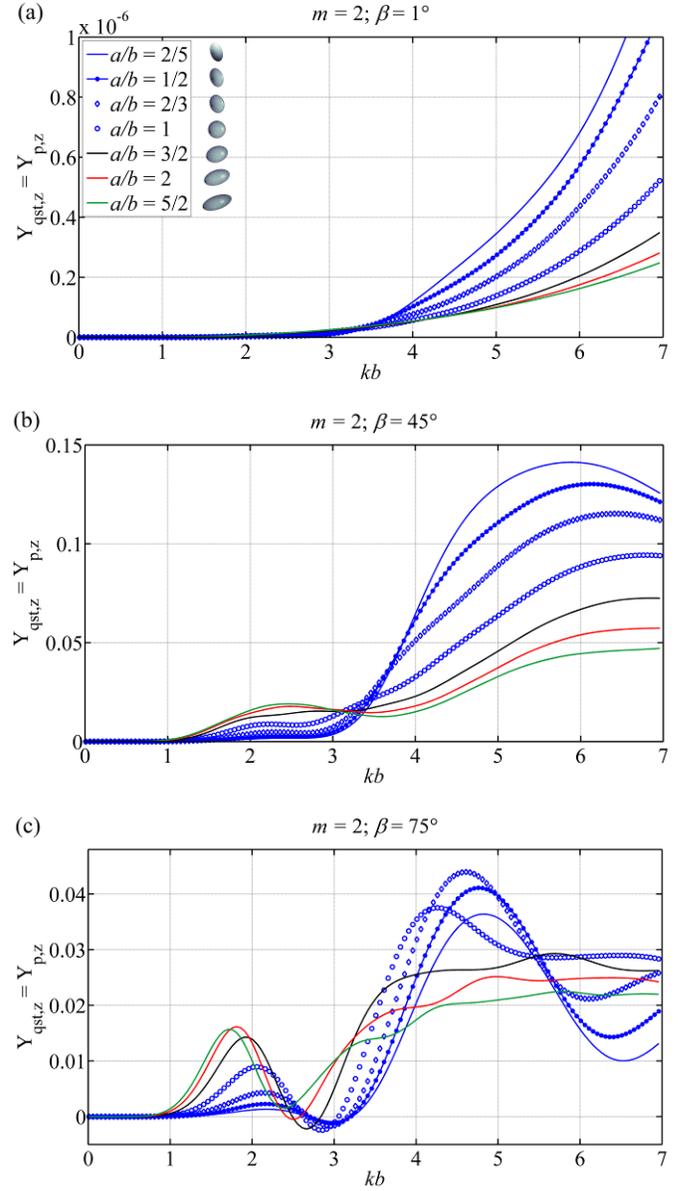

**Fig. 5.** The same as in FIG. 3, but the Bessel vortex beam is composed of (equi-amplitude) *standing waves* with $R = 1$.

**Fig. 6.** Panels (a)-(c) show the effect of varying the half-cone angle $\beta$ of a second-order Bessel vortex beam of *progressive waves* (i.e. $R = 0$) on the radiation force function given by Eq.(14) for rigid immovable (fixed) spheroids with end-on incidence.

The transition from the progressive to the *quasi-standing waves* case for a first-order Bessel vortex beam is shown in panels (a)-(c) of Fig. 4, with $R = 0.5$. Clearly, the effect of increasing the half-cone angle has a significant effect on the axial radiation force function for quasi-standing waves $Y_{qst,z}$, which is larger (in absolute sense) the more prolate the spheroid becomes. Moreover, the plots in Fig. 4 oscillate between positive as well as negative values while $kb$ varies, and depending on the position of the spheroid in the quasi-standing wave field. The force on the spheroid is directed towards a pressure node when $Y_{qst,z} > 0$, and to a pressure antinode when $Y_{qst,z} < 0$. The spheroid immobilization in the quasi-standing wave-field occurs at the equilibrium position occurs when $Y_{qst,z} = 0$. In addition, as $\beta$ increases to reach 75°, $Y_{qst,z}$ becomes $> 0$ for $kb < 2$ and for the chosen aspect ratios of the spheroid, which is pushed to a pressure node.

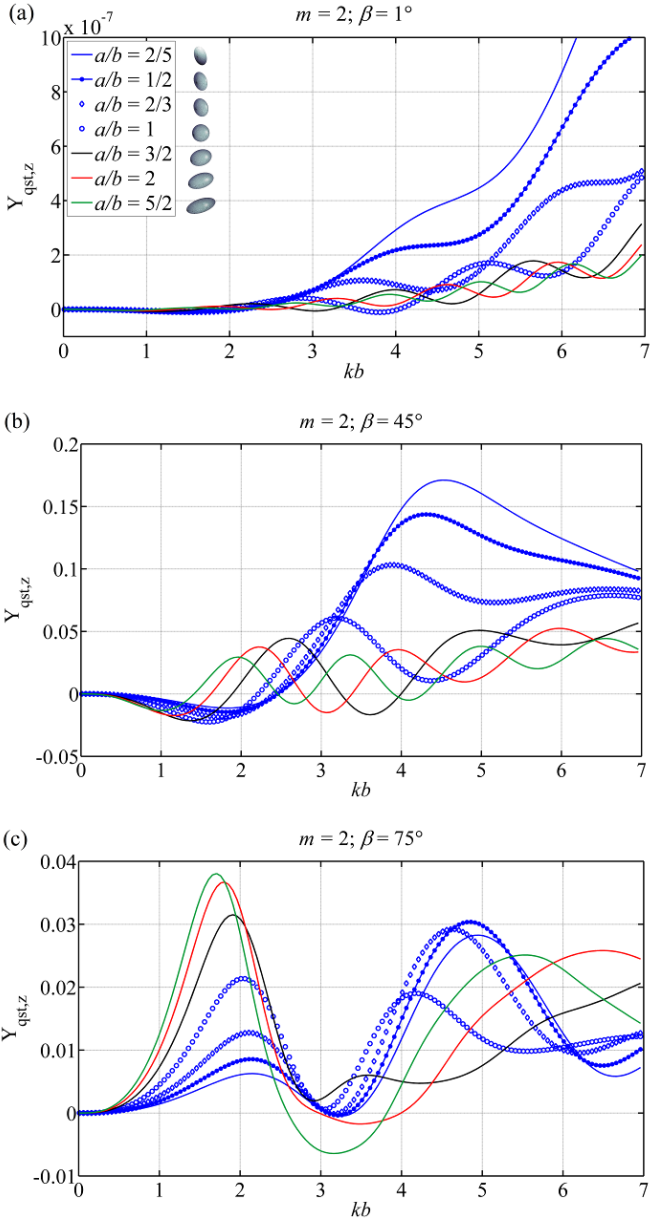
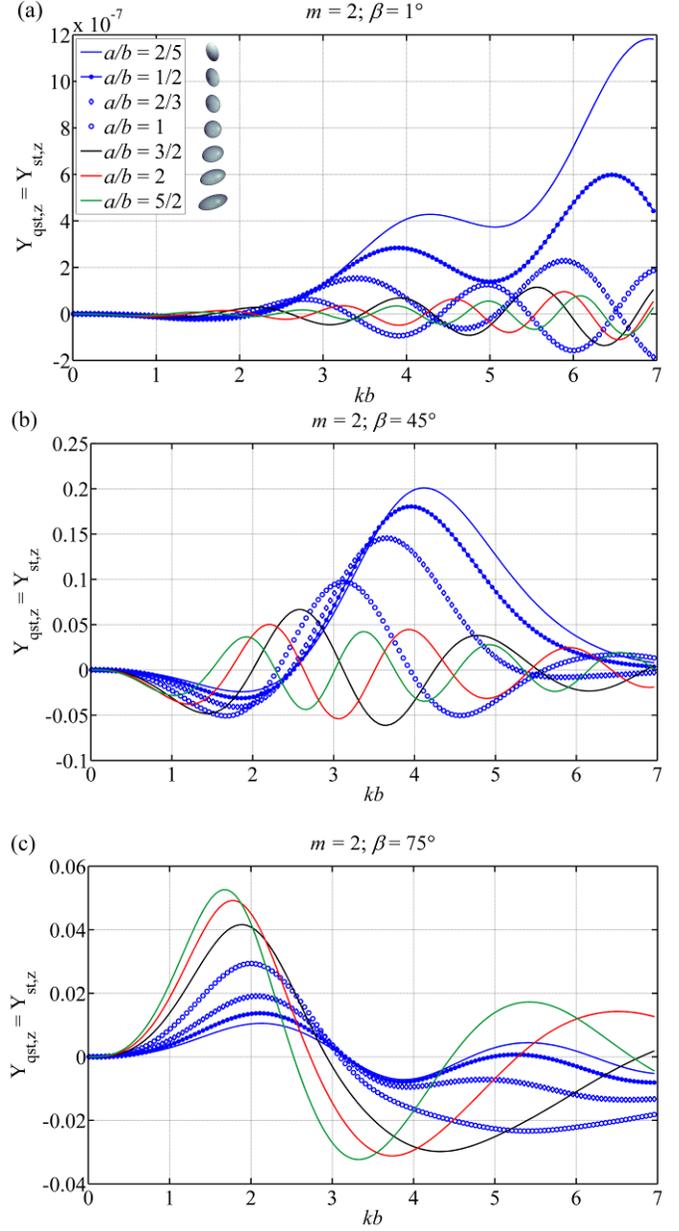

**Fig. 7.** The same as in Fig. 6, but the Bessel vortex beam is composed of quasi-standing waves with $R = 0.5$.

**Fig. 8.** The same as in Fig. 6, but the Bessel vortex beam is composed of equi-amplitude standing waves with $R = 1$.

A similar behavior in the plots for a first-order Bessel vortex beam of standing waves is observed from panels (a)-(c) of Fig. 5. The major differences, however, are manifested by an increase in the strength (in absolute sense) of the radiation force function plots for standing waves. This result may be anticipated from Eq.(14); this equation shows that for $0 < |R| < 1$, the function $Y_{qst,z}$ generally results from the contribution of two components, one induced by a standing wave, and the other by a progressive wave. As $R$ increases from 0 to 1, the factor $\frac{1}{2}(1-R^2)$ decreases from ½ to 0, which may lead (depending also on the half-cone angle values) to an increase in the magnitude plots in the absolute sense.

The effect of increasing the order of the beam to $m = 2$ is shown in Figs. 6-8, for which the corresponding panels show in general a comparable behavior to the plots of Figs. 3-5, respectively. Nonetheless, a decrease in the plots' magnitudes [in the absolute sense for panels (b) and (c)] is observed, since a second-order Bessel beam would carry less intensity/energy than a first-order Bessel beam for the same half-cone angle value along the direction of wave propagation. Furthermore, the plots for the progressive wave case shown in Fig. 6 approach zero for $kb \lesssim 1$, whereas for the quasi-standing (Fig. 7) and equi-amplitude standing wave (Fig. 8) cases, the plots tend to zero for $kb \lesssim 0.5$.

## 4. Conclusions and perspectives

The results presented in this analysis are the first to demonstrate the transition from the progressive to the equi-amplitude standing wave behavior of the acoustic radiation force of high-order Bessel vortex beams on rigid spheroids. The single "tractor beam" behavior of Bessel vortex beams as well as the dual-beam tweezers effects have been rigorously verified and established for oblate and prolate spheroids, assuming end-on incidence. One main advantage of the present semi-analytical formalism is the ability to predict the acoustic radiation force with numerical simulations using the PWSE of spherical multipoles (ordinarily used for a spherical particle) by forcing the expansion of the scattered field by means of outgoing spherical wave functions to satisfy the Neumann boundary condition for a rigid immovable surface. The method, which requires a single angular integration, is analyzed with particular emphasis on the half-cone angle $\beta$ of the Bessel vortex beam, the dimensionless size parameter $kb$, the aspect ratio $a/b$ (= major axis/minor axis) of the spheroid, the order $m$ of the Bessel vortex beam as well as the reflection coefficient $R$ forming the waves. The present formalism may be utilized to validate results obtained by strictly numerical methods. In addition, potential applications include the numerical predictions of the acoustic radiation force of Bessel vortex beams in underwater acoustics, particle levitation and manipulation, fluid dynamics, and generally any application dealing with the interaction of acoustical Bessel vortex beams with spheroidal objects, or other irregularly-shaped objects, such as super-spheroids, Chebyshev particles, finite cylinders or other arbitrary-shaped surfaces.

When designing acousto-fluidic devices for particle manipulation, estimation of the axial acoustic radiation force is used primarily as *a priori* information for optimization and improved performance purposes. For example, in a trapping system, the dimensions of the device are tweaked according to the force strength in order to retain a particle (or many particles) against a flow at a fixed position in a channel. Subsequently, the numerical/computational predictions of the force become an integral tool during the developmental process as well as the understanding of complex acousto-fluidic phenomena [2]. As such, when dealing with a spheroidal particle in Bessel beams of progressive, standing or quasi-standing waves (or even plane waves), the results presented here would be very beneficial since they anticipate adequate quantitative measures of the radiation force that could be used to advantage in optimized acousto-fluidic design. Apart from the axial component of the radiation force, the transverse (i.e. off-axial) components [13] also play an important role in particle immobilization and the stable positioning in acoustical potential wells and traps. Furthermore, another type of force that arises when multiple particles are used is the interparticle force [62]. It is created from multiple scattering effects between the particles being trapped so that attraction among the scatterers may stabilize the trapped cluster. The present analysis should assist in the development of improved analytical formalisms for spheroidal particles along those lines of researches.